\documentclass[12pt]{article}
\usepackage{graphicx}
\usepackage{lineno} 

\newcommand\pubnumber{SNSN-323-63}
\newcommand\pubdate{\today}

\def\institute{Physics Division\\
Lawrence Berkeley National Lab., Berkeley, CA 94720, USA}
\def\support{\footnote{Work supported by the Office of Science, Office of High Energy Physics, of the U.S.
Department of Energy under contract DE-AC02-05CH11231.}}
\def\copyright{\footnote{Copyright [2019] CERN for the benefit of the ATLAS Collaboration. CC-BY-4.0 license.}}

\def\Title#1{\begin{center} {\Large #1 } \end{center}}
\def\Author#1{\begin{center}{ \sc #1} \end{center}}
\def\Address#1{\begin{center}{ \it #1} \end{center}}

\newcommand\pubblock{\rightline{\begin{tabular}{l} \pubnumber\\
         \pubdate  \end{tabular}}}
\newenvironment{Abstract}{\begin{quotation}  }{\end{quotation}}
\newenvironment{Presented}{\begin{quotation} \begin{center} 
             PRESENTED AT\end{center}\bigskip 
      \begin{center}\begin{large}}{\end{large}\end{center} \end{quotation}}

\def\beq{\begin{equation}}
\def\eeq#1{\label{#1}\end{equation}}
\def\eeqn{\end{equation}}

\def\beqa{\begin{eqnarray}}
\def\eeqa#1{\label{#1}\end{eqnarray}}
\def\eeqan{\end{eqnarray}}

\let\bar=\overbar

\def\Dslash{\not{\hbox{\kern-4pt $D$}}}
\def\dslash{\not{\hbox{\kern-2pt $\del$}}}

\def\msb{{\bar{\ssstyle M \kern -1pt S}}}

\begin{document}
\begin{titlepage}
\pubblock

\vfill
\Title{Higgs(general) at ATLAS}
\vfill
\Author{ Wei-Ming Yao\support$^,$\copyright \\ On behalf of the ATLAS collaboration}
\Address{\institute}
\vfill
\begin{Abstract}
The ATLAS Higgs results are reviewed using Run-2 data taken at a center-of-mass energy of 13 TeV with up to 
an integrated luminosity of 80 fb$^{-1}$. So far, the data are consistent with the standard model expectations. 
ATLAS now has observed the Higgs Yukawa coupling to 
the third generation fermions with $H\rightarrow \tau\tau$, ttH, and $H\rightarrow b b$ in the VH process. 
The Higgs boson will continue to provide an important probe for new physics and beyond.   
\end{Abstract}
\vfill
\begin{Presented}
$11^\mathrm{th}$ International Workshop on Top Quark Physics\\
Bad Neuenahr, Germany, September 16--21, 2018
\end{Presented}
\vfill
\end{titlepage}
\def\thefootnote{\fnsymbol{footnote}}
\setcounter{footnote}{0}

\section{Introduction}

The observation of Higgs boson (H) by ATLAS and CMS in July 2012~\cite{hdiscoveryatlas,hdiscoverycms} marked the completion of
the standard model (SM) and opened the way to explore the Higgs sector that is
responsible for electroweak symmetry breaking (EWSB). 
In this contribution, I will review some recent 
ATLAS Higgs results using Run-2 data at a center-of-mass energy of 13 TeV with up to an integrated luminosity of 80 fb$^{-1}$.

\section{Higgs production and property measurements}

The Higgs boson at the LHC is predominantly produced via the ggF, VBF, VH, and ttH processes. 
There are five H decay channels playing an important role at the LHC, which are $H\rightarrow b\bar b$, $WW^*$, $\tau\tau$, $ZZ^*$, and 
$\gamma\gamma$. 

\subsection{Higgs bosonic decays}
\begin{figure}[htb]
\centering
\includegraphics[height=1.5in]{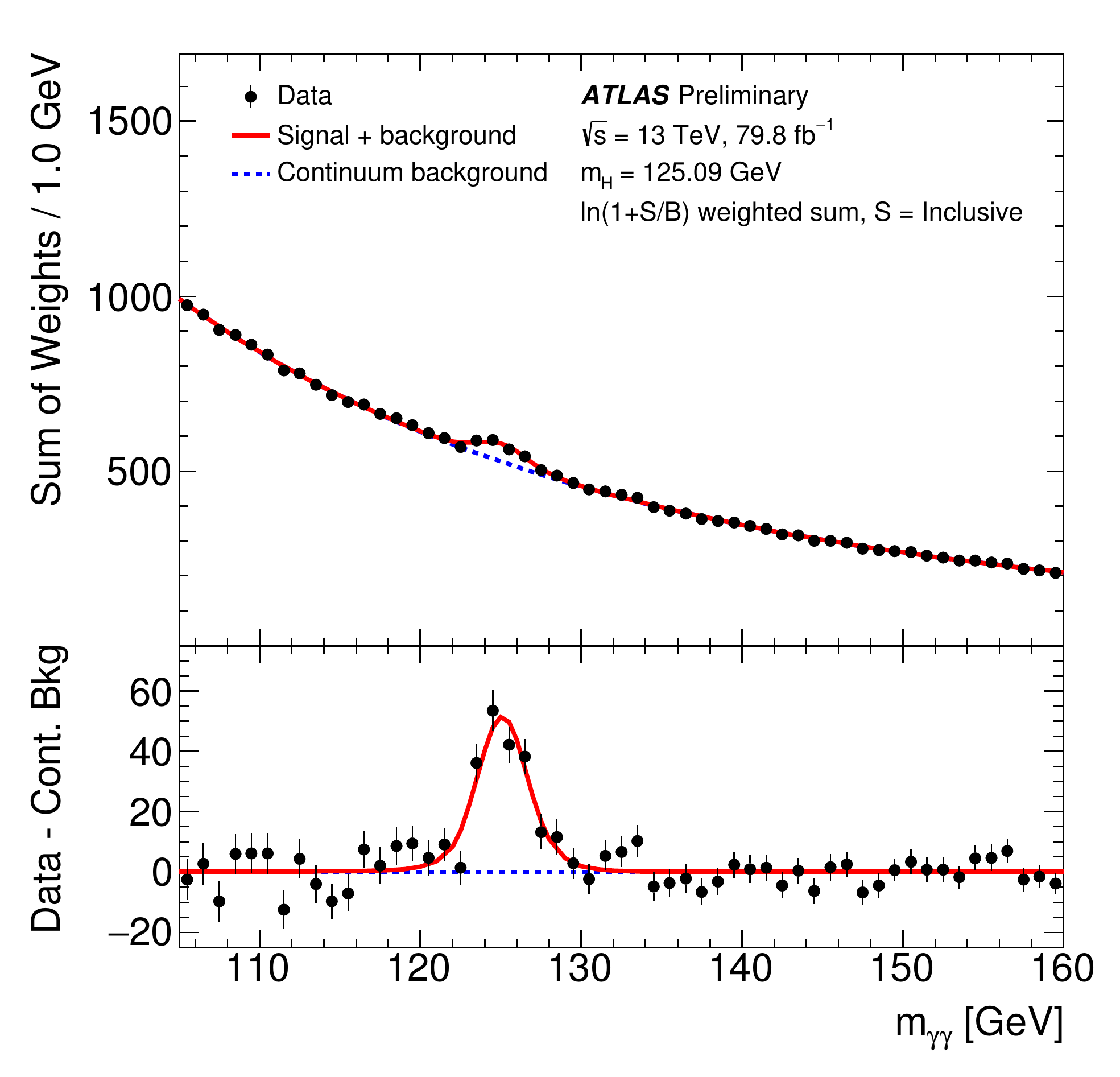}
\includegraphics[height=1.5in]{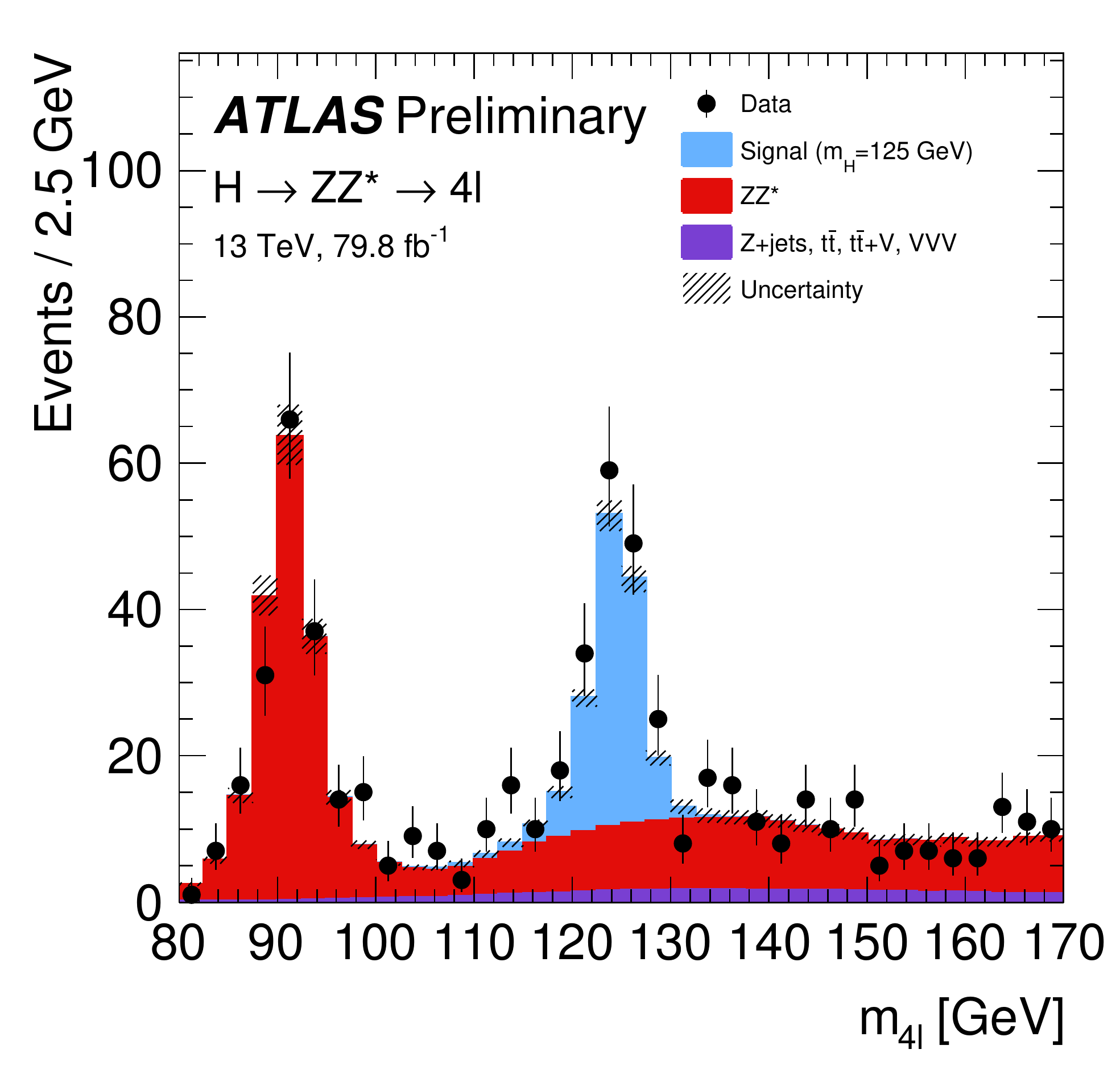}
\includegraphics[height=1.5in]{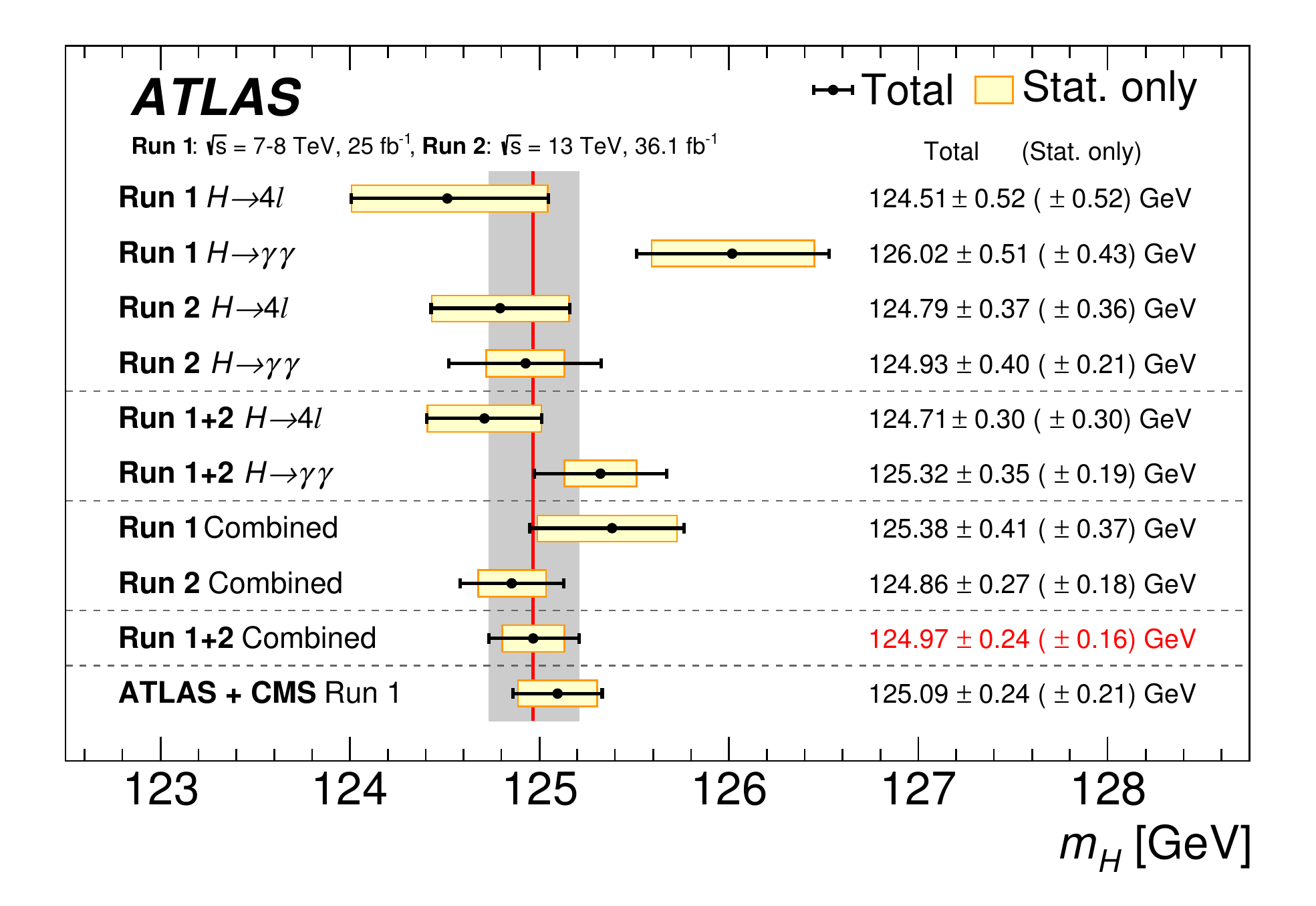}
\caption{The weighted diphoton invariant mass spectrum in all the analysis categories(left) and the inclusive four-lepton invariant mass(middle) 
are observed in 79.8 fb$^{-1}$ of 13 TeV data. The summary of $m_H$ measurements from the individual and combined analyses (right) use 
36 fb$^{-1}$ of Run-2 data~\cite{run2gammagamma, run24l, LHCmass}.}  
\label{fig:Hmgg4l}
\end{figure}

  The Higgs production is studied in the $H\rightarrow \gamma\gamma$ and $H\rightarrow 
ZZ^*\rightarrow 4l$ decay channels using Run-2 data~\cite{run2gammagamma, run24l}. Figure~\ref{fig:Hmgg4l} 
shows the invariant masses of di-photon and the four-leptons. The signal strength 
is defined as the ratio of the observed yield to its SM prediction. It is measured to be 
$\mu_{H\rightarrow \gamma\gamma}=1.06\pm 0.08^{+0.11}_{-0.09}$ and $\mu_{H\rightarrow 4l}=1.19\pm 0.12^{+0.10}_{-0.09}$, respectively. 
Many fiducial and differential cross sections are also measured,
 which are in excellent agreement with the Monte Carlo predictions.
 
   The $H\rightarrow \gamma\gamma$ and $H\rightarrow 4l$ events are also used to measure $m_H$ in Run-2 data~\cite{run2mass}. 
The results from each of the individual channels and their combination with Run-1 are shown in Figure~\ref{fig:Hmgg4l}.
The ATLAS combined Higgs mass is $m_H=124.97\pm 0.24$ GeV, which is compatible with the LHC Run1 measurement of $125.09\pm 0.24$ GeV~\cite{LHCmass}. 

  The Higgs boson width ($\Gamma_H$) is expected to be too small to be measured directly at the LHC,
but it can be constrained indirectly through the off-shell $H\rightarrow ZZ$ 
production and background interference effects~\cite{Hoffshelleffects}.
Studies based on the channels $H\rightarrow ZZ^*\rightarrow 4l$ and $H\rightarrow ZZ\rightarrow 2l2\nu$
using Run2 data~\cite{Hwidth} allowed to set an upper limit, at 95\% confidence level, of 14.4~MeV (15.2~MeV expected)
on $\Gamma_H$.

\subsection{Higgs fermionic decays} 

The Higgs Yukawa coupling can be probed either indirectly at the loop
level or directly at the tree-level. Any deviation from SM could be a sign of new physics.
The discussion of the ATLAS observation of ttH can be found elsewhere~\cite{ttH}. 

\begin{figure}[htb]
\centering
\includegraphics[height=1.5in]{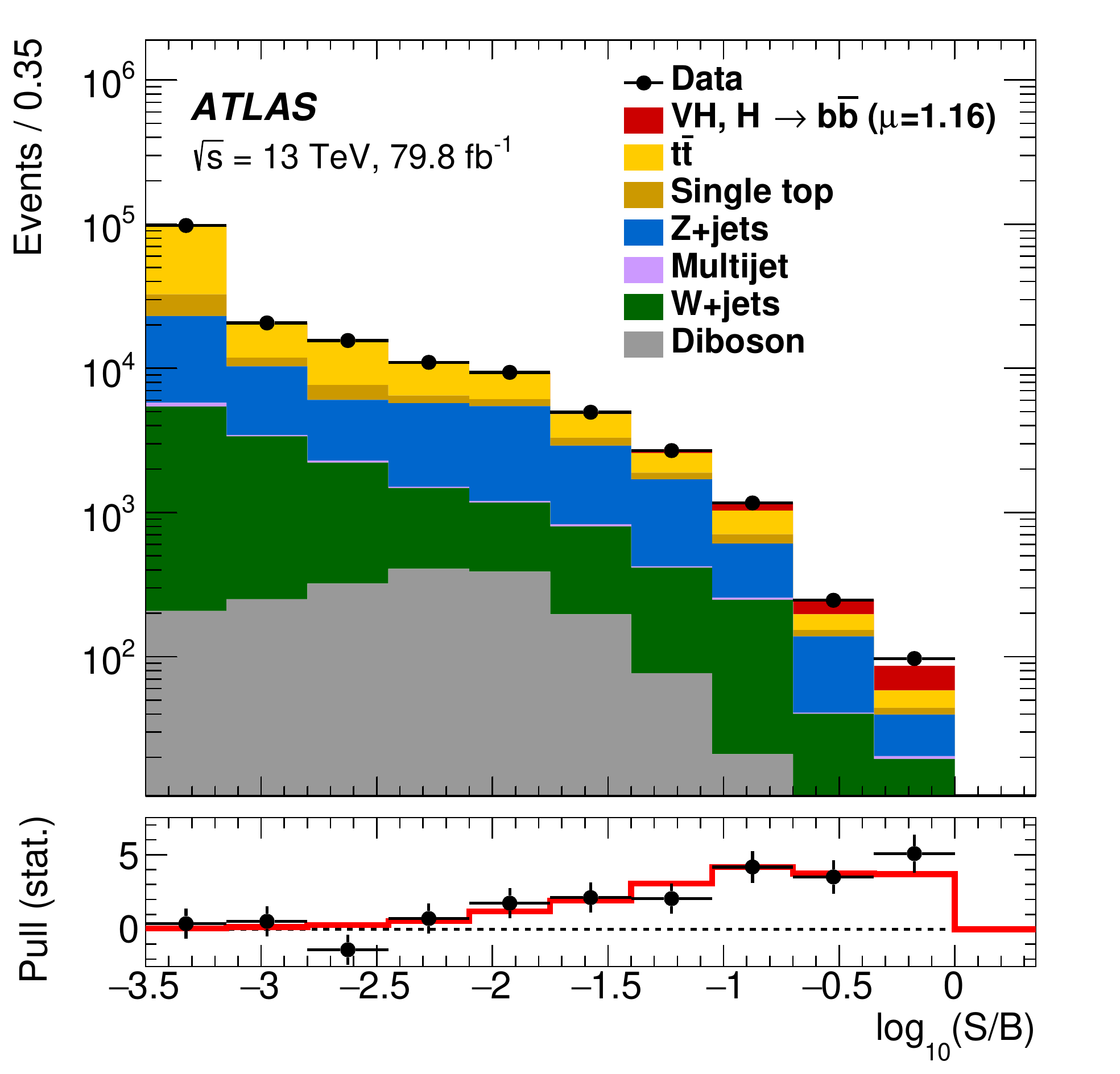}
\includegraphics[height=1.5in]{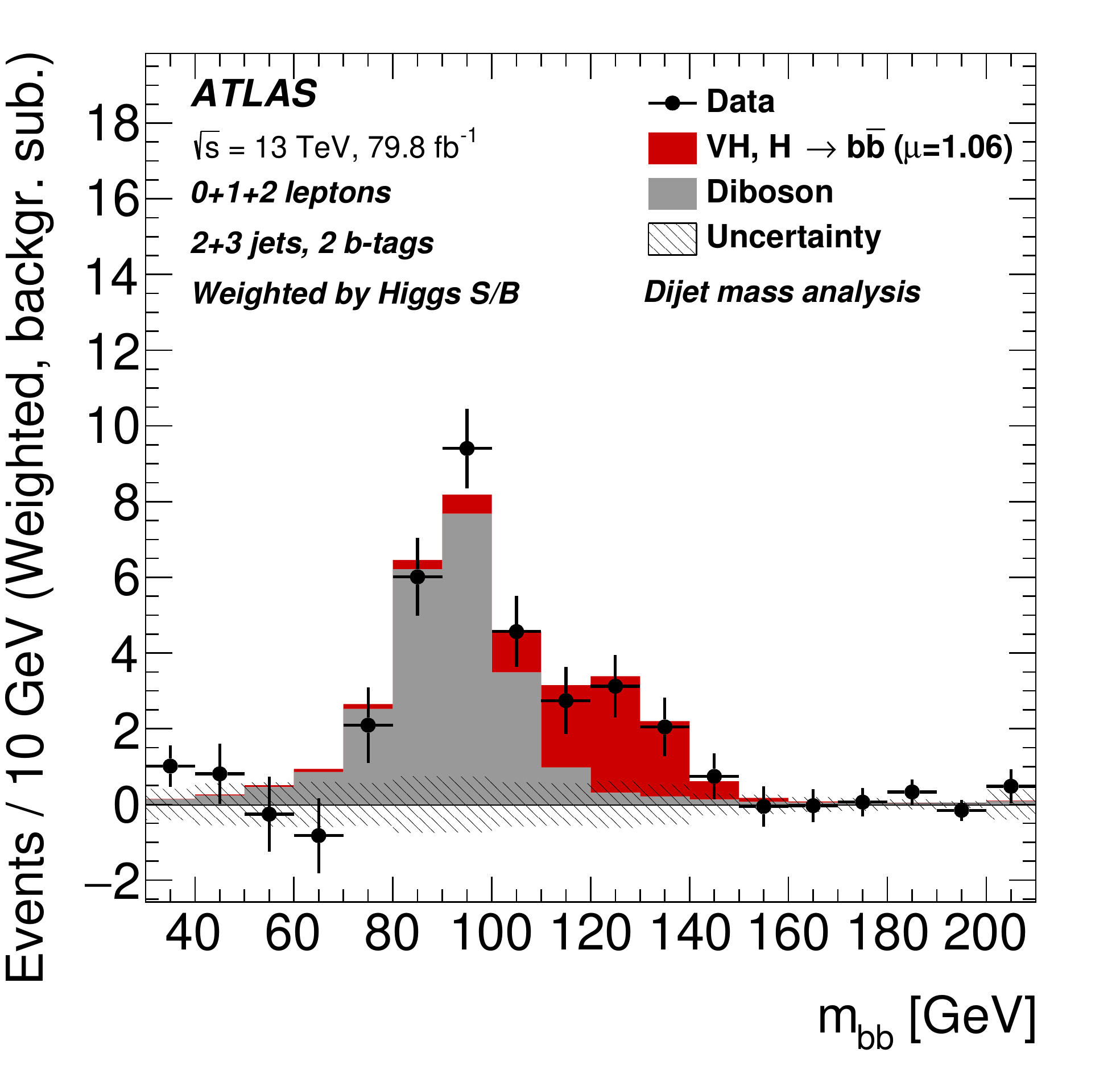}
\includegraphics[height=1.5in]{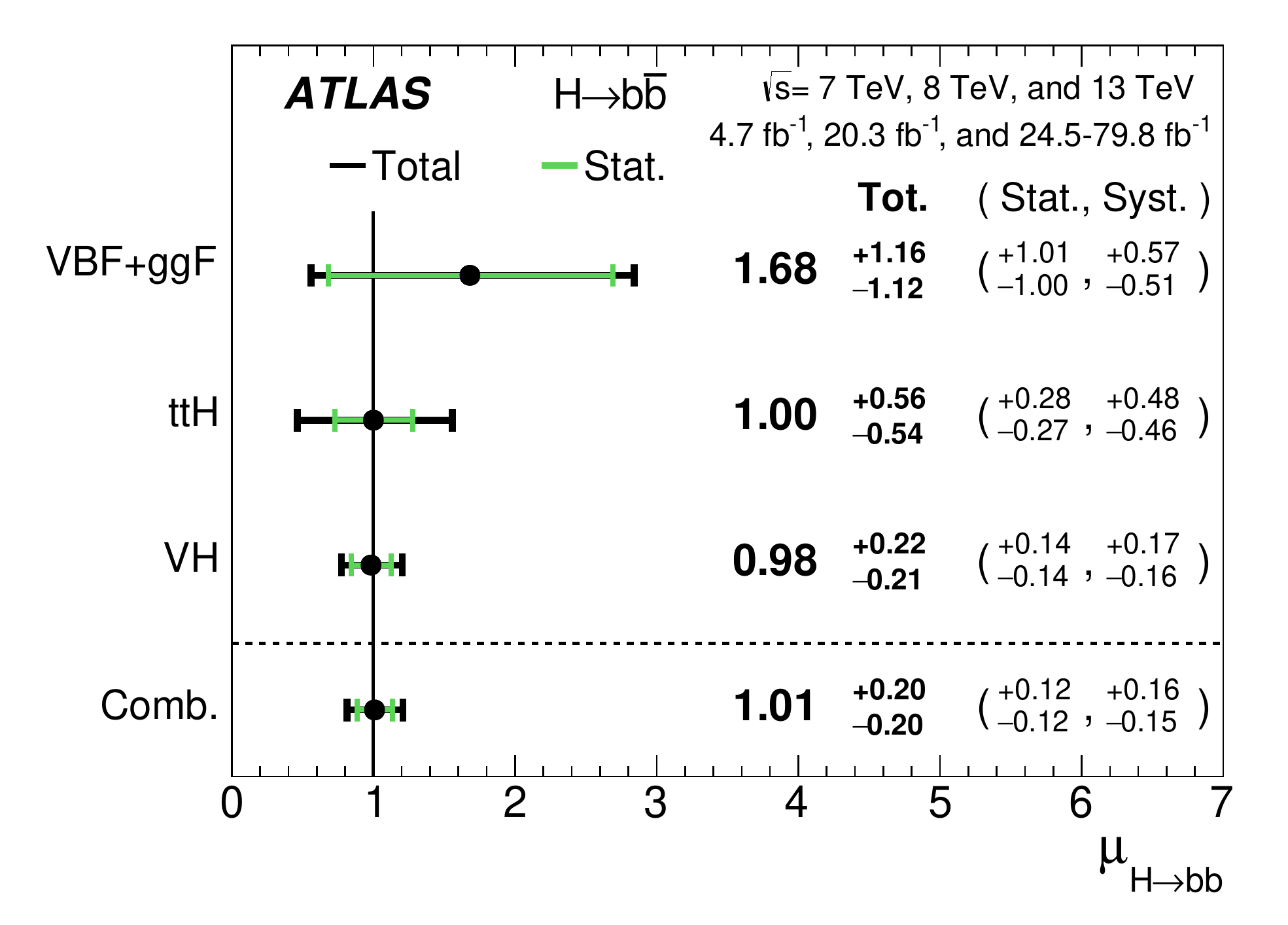}
\caption{In the search for VH(bb),event yield as a function of log(S/B) (left) and the $m_{bb}$ in data after subtraction of all background with the 
dijet-mass analysis (middle) are shown in 79.8 fb$^{-1}$ of Run2 data. Fitted $\mu_{H\rightarrow b b}$ 
from VH, ttH, and VBF+ggF along with their combinations using all the data (right)~\cite{Run2VH}.}
\label{fig:Hbb}
\end{figure}

ATLAS has recently observed $VH$ production ($V=W$ or $Z$) with $H\rightarrow b\bar b$~\cite{Run2VH}.
The VH process is most sensitive to $H\rightarrow b\bar b$.
In order to separate signal from background, the analysis is divided into several channels based on the number of charged 
leptons from the V decay and a trained BDT. 
Figure~\ref{fig:Hbb} shows a cumulative log(S/B) from all channels.
The fitted signal strength is $\mu_{VH\rightarrow b b}=1.16 \pm 0.16\pm 0.20$, which
gives an observed significance of 4.9~$\sigma$ with an expected of 4.3~$\sigma$.
The combination with Run-1 data gives $\mu_{VH\rightarrow b b} =0.98\pm0.14\pm0.16$ and
an observed significance of 4.9~$\sigma$ (expected 5.1~$\sigma$).
We have also repeated the search by fitting the $b\bar b$ invariant mass only as shown in Figure~\ref{fig:Hbb}, 
giving a consistent result. ATLAS further combines the $H\rightarrow b\bar b$ results in VH 
with other processes of VBF and ttH, resulting in 
observed significance of 5.4~$\sigma$ (expected 5.5~$\sigma$).
The VH production is also combined using three H decay channels $b\bar b$, $\gamma\gamma$, and $4l$,
which gives an observed significance of 5.3~$\sigma$ (expected 4.8~$\sigma$).

The $H\rightarrow \tau\tau$ channel has results from Run-2 with 36.1 fb$^{-1}$~\cite{Htautau}.
The di~-~$\tau$ events are selected in three final states of  $\tau_{lep} \tau_{lep}$, $\tau_{lep} \tau_{had}$, and $\tau_{had} \tau_{had}$
that are further divided into 13 categories targeting VBF production and ggF in the boosted region. 
The signal strength is measured to be $\mu_{H\rightarrow \tau\tau}=1.09{^{+0.18}_{-0.17}}{^{+0.31}_{-0.25}}$, which gives an 
observed significance of 4.4~$\sigma$ (expected 4.1~$\sigma$).
After combining with Run-1 results, an observed significance of 6.4~$\sigma$ (expected 5.4~$\sigma$) is obtained. 
The 2-D fitted ggH and VBF cross sections are in good agreement with the SM predictions.

The search for the $H\rightarrow \mu\mu$ decay is also updated using 80 fb$^{-1}$ of Run-2 data~\cite{Hmumu}.
No excess of events over the known backgrounds is observed and a 95\% CL upper limit 
on the signal strength of 2.1 (expected 2.0) is set.

\subsection{Higgs boson couplings} 
 
\begin{figure}[htb]
\centering
\includegraphics[height=1.5in]{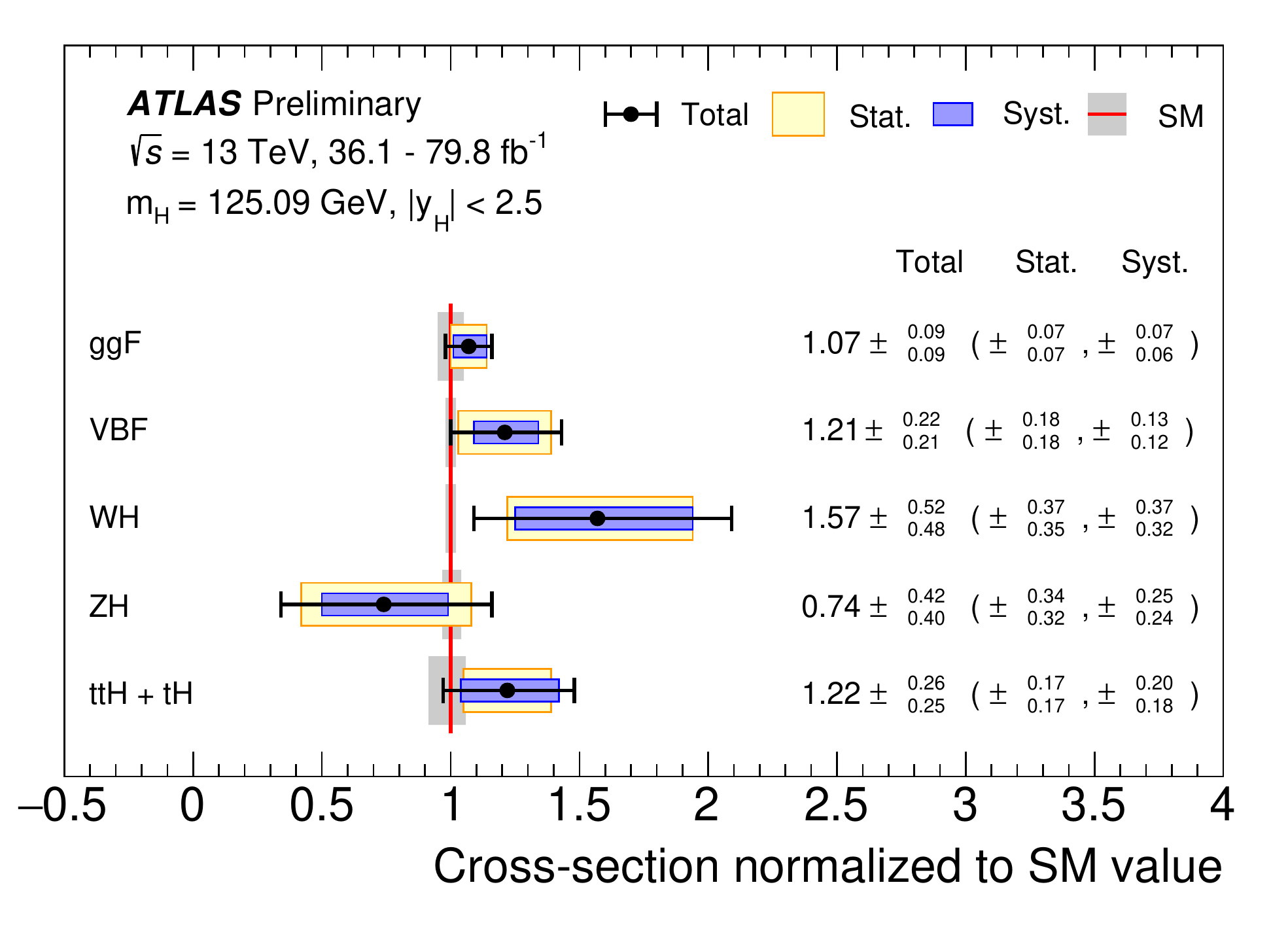}
\includegraphics[height=1.5in]{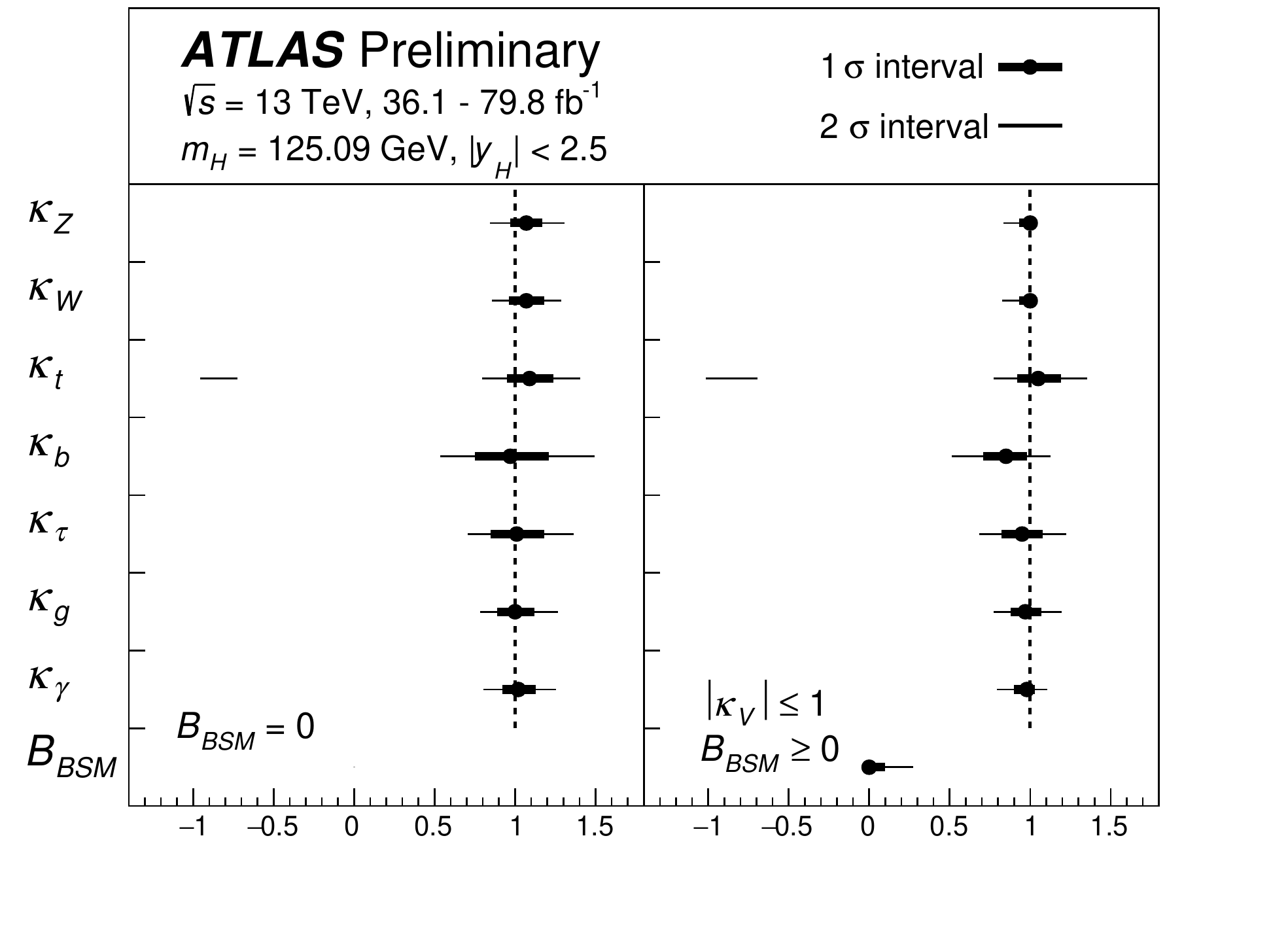}
\includegraphics[height=1.5in]{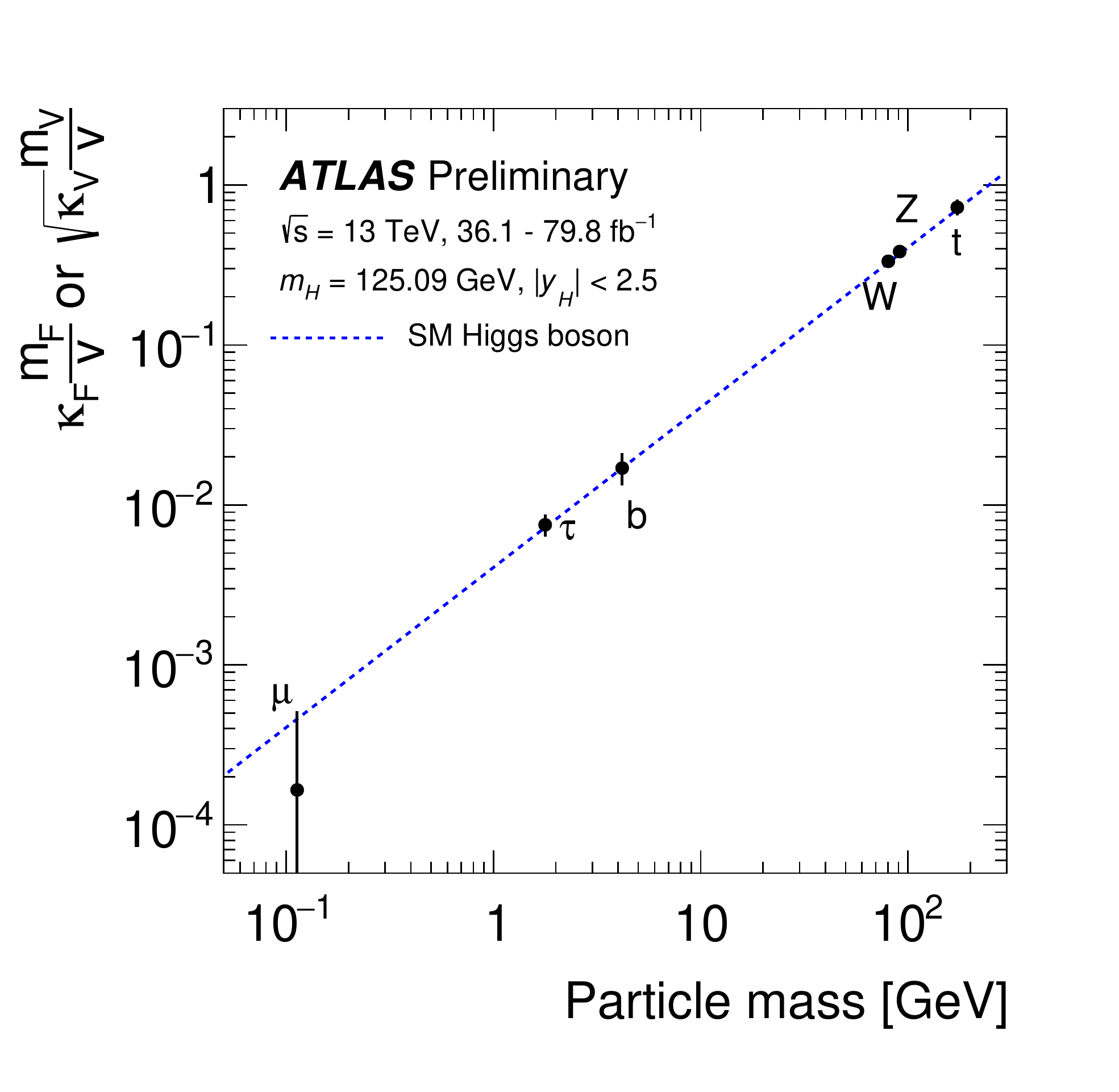}
\caption{The measured Higgs signal strength for ggF, VBF, VH, and ttH+tH (left), the best-fit of H coupling modifiers per particle type 
by either $B_{BSM}=0$ or $B_{BSM}$ included as a free parameters (middle), and the Higgs coupling for fermions and bosons as a function of their
masses (right) are shown~\cite{combinedfit}.}
\label{fig:Hcombined}
\end{figure}

Since the main Higgs production modes have been observed by ATLAS, the combined results are used to check the consistency of the 
couplings with the SM and to probe
BSM coupling as kappa modifiers~\cite{combinedfit}. The combined signal strengths for ggF, VBF, VH, and ttH+tH are shown in Figure~\ref{fig:Hcombined}. 
Two tests are performed and shown. One test assumes no physics beyond the Standard Model; in this case kappa modifiers~\cite{combinedfit} of the couplings are 
left floating in the fit to data simultaneously. The result is consistent with 
the SM with a CL of 87\%. In a second test, the parameter $B_{BSM}$ is added to the kappa-modifiers to describe explicit contributions to the Higgs
width from new physics. The result, $B_{BSM}<0.26$ (expected 0.37) at 95\% CL, is consistent with direct searches based on Run-1 data~\cite{Run1invisible}.
Figure~\ref{fig:Hcombined} also shows
the Higgs coupling to fermions and bosons as a function of their masses, which are in good 
agreement with the SM expectation. 

\section{Search for di-Higgs production}

Probing Higgs self-coupling at LHC is extremely challenging.
ATLAS has searched SM di-Higgs production using 36 fb$^{-1}$ of Run-2 data with three major di-Higgs decay
modes: $HH\rightarrow 4b$, $HH\rightarrow bb\tau\tau$, $HH\rightarrow bb\gamma\gamma$, 
and their combination~\cite{dihiggs}.
No excess of events is observed and the combined limit on the signal strength is set less than 6.7 (10.4 expected) at 95\% CL. 
The corresponding observed limit at 95\% CL on  Higgs self-coupling is between -5.0 and 12.1 (expected limit
is between -5.8 and 12.0). 

\section{Conclusion}
In conclusion, the ATLAS Higgs results are reviewed using Run-2 data taken at a center-of-mass energy of 13 TeV with up to
an integrated luminosity of 80 fb$^{-1}$. So far, the data are consistent with the standard model expectations.
ATLAS now has observed $H\rightarrow \tau\tau$, ttH, and $VH\rightarrow b b$.
The Higgs boson will continue to provide an important probe for new physics and beyond.



\end{document}